\documentclass[twocolumn,pre,showpacs,floatfix]{revtex4}
\usepackage{graphicx} \usepackage{bm}
\newcommand{\be}{\begin{equation}}
\newcommand{\ee}{\end{equation}}

\begin{document}
\title{Steady state solutions of hydrodynamic traffic models}
\author{H. K. Lee,$^1$ H.-W. Lee,$^2$ and D. Kim$^1$}
\affiliation{ $^1$School of Physics, Seoul National University,
Seoul 151-747, Korea } \affiliation{$^2$Department of Physics,
Pohang University of Science and Technology,
 Pohang, Kyungbuk 790-784, Korea }
\date{\today}

\begin{abstract}
We investigate steady state solutions of hydrodynamic traffic
models in the absence of any intrinsic inhomogeneity on roads such
as on-ramps. It is shown that typical hydrodynamic models possess
seven different types of inhomogeneous steady state solutions. The
seven solutions include those that have been reported previously
only for microscopic models. The characteristic properties of wide
jam such as moving velocity of its spatiotemporal pattern and/or
out-flux from wide jam are shown to be uniquely determined and
thus independent of initial conditions of dynamic evolution.
Topological considerations suggest that all of the solutions
should be common to a wide class of traffic models. The results
are discussed in connection with the universality conjecture for
traffic models. Also the prevalence of the limit-cycle solution in
a recent study of a microscopic model is explained in this
approach.
\end{abstract}
\maketitle

\section{Introduction}
Vehicles on roads interact with other and various traffic
phenomena can be regarded as collective behaviors of interacting
vehicles~\cite{Nagel92JP,Kerner93PRE,Bando95PRE}. Analyses of
highway traffic data revealed that there exist qualitatively
different states of traffic
flow~\cite{Treiterer74Proceeding,Kerner96PRE,Neubert99PRE,Treiber00PRE,Lee00PRE}.
Transition between different traffic states was also
reported~\cite{Kerner96PRE}. Some empirical findings, such as the
so-called synchronized traffic flow phase~\cite{Kerner96PRE}, have
become subjects of considerable concern and ignited intense
theoretical investigations of traffic flow. Comprehensive reviews
can be found for example in Ref.~\cite{Chowdhury00PR}.

Many traffic flow models have been put forward and
analyzed~\cite{Komatsu95PRE,Nagatani97JPSJ,Lee98PRL,Helbing98PRL,Helbing99PRL,
Lee99PRE,Mitarai99JPSJ,Knospe00JPA,
Tomer00PRL,Nelson00PRE,Lubashevsky00PRE,Safanov02EP}. Quite often,
the analysis aims to find out possible steady states, or dynamic
phases, of the models and investigate their properties. A
criterion for good and reliable traffic models is good agreement
between steady states of models and traffic states revealed via
real traffic data analysis. Thus an important first step in the
analysis of traffic flow models is to find out all possible steady
states.

In this paper, we consider hydrodynamic models for traffic without
bottlenecks such as ramps
and present a systematic search for their steady state solutions
which are time-independent in a proper moving reference frame,
\begin{equation}
   \rho(x,t) = \tilde \rho(x + v_g t), \ \
   v(x,t) = \tilde v(x + v_g t),
   \label{final_Field}
\end{equation}
where $\rho$ and $v$ are density and velocity fields,
respectively, and $-v_g$ is the constant velocity of the moving
reference frame with respect to the stationary reference frame.
Two well-known steady solutions of this type are free flow and
traffic jam solutions~\cite{Kerner94PRE,Wada97CM,Kerner97PRE}.
Surprisingly, we find that hydrodynamic models possess not only
these two but also {\it several} other steady state solutions.
Some of the newly recognized steady state solutions, including
limit cycle solutions, have been reported previously only for
microscopic traffic
models~\cite{Mitarai99JPSJ,Tomer00PRL,Berg01PRE} and not been
reported for hydrodynamic models, which led to the wide spread
view that free flow and traffic jam are the only possible steady
state solutions of hydrodynamic models in the absence of
bottlenecks such as ramps. Our result shows that such view is
incorrect and that the physics contained in hydrodynamic models
may be much richer than previously recognized.

In Sec.~\ref{ssov}, we first review the mapping to the single
particle motion and introduce the concept of the flow diagram in
the single particle phase space. It is demonstrated in
Sec.~\ref{flow-diagram-structures} that the flow diagram can have
various topologically different structures, which are directly
linked to the existence of certain types of steady state solutions
(Sec.~\ref{steady-state-solutions}). Section~\ref{discussion}
discusses implications of our results. Section~\ref{conclusion}
concludes the paper.

\section{Mapping to single particle motion}
\label{ssov}
We consider a hydrodynamic model that
consists of the following two equations,
an equation for local vehicle number conservation,
\begin{equation}
   {\partial\over{\partial t}}\rho(x,t) +
   {\partial\over{\partial x}} \left [\rho(x,t)v(x,t)\right ] = 0,
   \label{VehicleConservation}
\end{equation}
and an equation of motion,
\begin{equation}
    {{\partial v} \over {\partial t}} +
   v{{\partial v}\over{\partial x}} =
   {\cal R} \left[V_{\rm op}\left(\rho^{-1}\right)
           - v\right]
        - {\cal A}
         {{\partial \rho}\over{\partial x}}
        +{\cal D}
                {{\partial^2 v}\over{{\partial x}^2}},
   \label{VelocityTotalDeriv_Final}
\end{equation}
where $V_{\rm op}\left(\rho^{-1}\right)$ is the so-called optimal
velocity function and the coefficients (${\cal R}$ for the
relaxation term, ${\cal A}$ for the anticipation term, and ${\cal
D}$ for the diffusion term) are positive definite and depend in
general on the density and velocity fields, i.e., ${\cal
R}(\rho,v),~{\cal A}(\rho,v)$, and ${\cal D}(\rho,v)>0$.

To find out steady state solutions of the type in
Eq.~(\ref{final_Field}),
it is useful to map the problem into a single particle motion
problem by using the method in Refs.~\cite{Kerner94PRE,Kerner97PRE}.
For the mapping, one first integrates out
Eq.~(\ref{VehicleConservation}).
The resulting constant of motion,
\begin{equation}
   q_g = \rho \cdot (v + v_g),
   \label{Flux_in_vg}
\end{equation}
relates two dynamic fields $\rho$ and $v$,
and can be used to reduce the number of independent dynamic
fields from two to one (we choose $v$ in this work).
Then Eq.~(\ref{VelocityTotalDeriv_Final}) can be transformed into an
{\it ordinary} differential equation for the single dynamic
field $v$ that depends on a single parameter $z\equiv x+v_g t$,
\begin{equation}
   {{d^2 v}\over{dz^2}} + {\cal C}(v; v_g, q_g) {{dv}\over{dz}}
                                + {\cal F}(v; v_g, q_g) = 0,
   \label{Eqn_in_ss}
\end{equation}
where
\begin{eqnarray} \label{C&F}
{\cal C}(v; v_g, q_g) &\equiv& {1\over {\cal D}}
 \left[ { {\cal A}q_g \over (v+v_g)^2 } -(v+v_g) \right],
\\
{\cal F}(v; v_g, q_g) &\equiv& {{\cal R} \over {\cal D}}
 (V_{\rm op}-v). \nonumber 
\end{eqnarray}
Here the field $\rho$ in the arguments of ${\cal R},~{\cal
A},~{\cal D}$, and $V_{\rm op}$ should be understood as
$q_g/(v+v_g)$. Then the search for steady state solutions reduces
to the analysis of Eq.~(\ref{Eqn_in_ss}) under the physically
meaningful boundary condition that solutions should be bounded as
$z\to\pm\infty$.

\begin{figure}
\includegraphics[width=8.0cm, height=6.0cm]{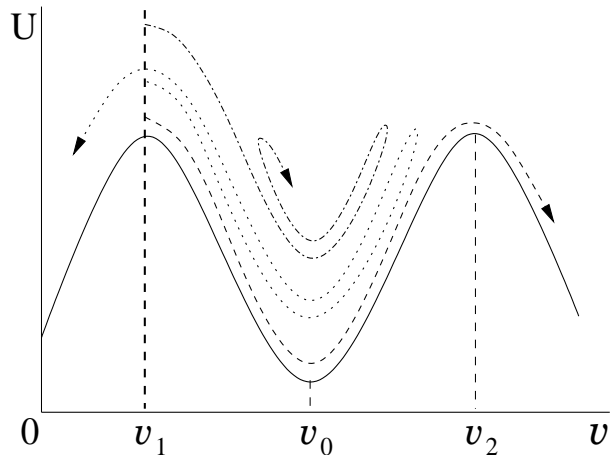}
\caption{A typical shape of the potential energy $U(v;v_g,q_g)$ (solid line).
Three examples of the particle motion are also shown (dotted, dashed,
dash-dotted lines).
}
\label{Lall}
\end{figure}

To gain insights into implications of Eq.~(\ref{Eqn_in_ss}), it is
useful to make an analogy with a classical mechanics of a particle
by regarding $z$ as a time variable and $v$ as a {\it coordinate}
of a particle with unit mass moving in a one-dimensional system.
Then Eq.~(\ref{Eqn_in_ss}) describes the time evolution of a
particle under the influence of a potential energy, $U(v; v_g,
q_g) = \int^{v} dv' {\cal F}(v'; v_g, q_g)$, and of a damping
force with the coordinate-dependent damping coefficient ${\cal
C}(v; v_g, q_g)$. For a physical choice of $V_{\rm
op}(\rho^{-1})$, which decreases as $\rho$ increases, goes to zero
for large $\rho$, and saturates at a finite value for small
$\rho$, the potential energy $U$ becomes {\it camelback-shaped}
for wide ranges of $v_g$ and $q_g$ (solid curve in
Fig.~\ref{Lall}). Thus the potential energy profile is of very
typical shape. Below we focus on $U$ of this shape only and ignore
the possibility of more exotic shaped $U$'s, such as $U$'s with
three peaks, since we do not know of any reason to expect such
exotic possibilities. Furthermore we will not address the trivial
case occurring when the range of $v_g$ and $q_g$ allows less than
two peaks in $U$.

An unusual feature in this mechanical analogy is that
the damping coefficient $\cal C$ is
{\it not} necessarily positive and
in those ranges of $v$, where $\cal C$ is negative,
the particle may {\it gain} energy due to the damping.
The possibility of the negative damping
is crucial for the existence of certain steady state solutions presented in the
next section.

Before we close this section, we remark that the stability of a
solution in the single particle problem should not be identified
with the stability in the real traffic problem. We demonstrate
this point with trivial $z$-independent solutions. For the
camel-back-shaped potential in Fig.~\ref{Lall}, extremal points of
the potential become solutions. Thus there are three
$z$-independent solutions, $v=v_i$ ($i=0,1,2$ and $0\le v_1 < v_0
<v_2$), where $v_0$ is the coordinate of the local minimum and
$v_1$, $v_2$ are coordinates of the two local maxima or saddle
points. All  $v_i$'s satisfy $V_{\rm op}((v+v_g)/q_g)= v$ and thus
depend on $v_g$ and $q_g$. These $z$-independent solutions
correspond to the homogeneous traffic states $v(x,t)=v_{0,1,2}$.
From the shape of the potential energy, it is clear that in the
single particle problem, the two solutions $v(z)=v_{1,2}$ are
unstable with respect to small deviations and the other solution
$v(z)=v_0$ is stable (if ${\cal C}$ is positive near $v_0$). For
the real traffic problem [described by
Eqs.~(\ref{VehicleConservation}) and
(\ref{VelocityTotalDeriv_Final})], however, the solutions
$v(x,t)=v_{1,2}$ are (usually) {\it stable} with respect to small
deviations, and $v(x,t)=v_0$ is linearly unstable.

\section{Flow diagrams}
\label{flow-diagram-structures}

Besides the trivial static ($z$-independent) solutions,
there also exist dynamic ($z$-dependent) solutions,
some examples of which are shown in Fig.~\ref{Lall} (dotted, dashed, and
dash-dotted lines).
In the language of traffic flow, these dynamic solutions
correspond to {\it steady} but {\it inhomogeneous} traffic flow states.
For a systematic study of dynamic solutions,
it is useful to introduce a two-dimensional phase space $(v,w\equiv dv/dz)$,
where each trajectory in the phase space corresponds to a dynamic solution.
Then finding all solutions for given $v_g$ and $q_g$ is
equivalent to constructing a flow diagram in the phase space
for the given $v_g$ and $q_g$.

\begin{figure}
\includegraphics[width=8.0cm, height=5.0cm]{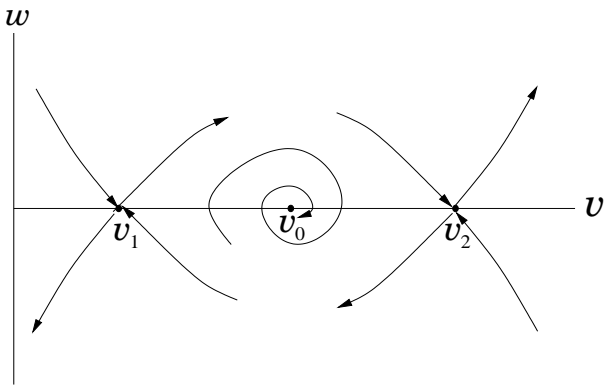}
\caption{A schematic diagram of the flow near the three fixed points
  $(v_{0,1,2},0)$. ${\cal C}$ is assumed to be positive near $v=v_0$
  in this figure.
}
\label{fixed-points}
\end{figure}

To gain an insight into flow diagram structures, it is useful to
examine the flow near the three fixed points $(v=v_i,w=0)$'s.
Figure~\ref{fixed-points} show flows near the three fixed points.
The fixed points $(v_{1,2},0)$ are saddle points regardless of the
sign of ${\cal C}$ near $v=v_{1,2}$ since the damping force alone
[the second term in Eq.~(\ref{Eqn_in_ss})] can not reverse the
direction of the particle motion (that is, the sign of $dv/dz$)
even if the damping coefficient is negative. On the other hand,
the fixed point $(v_0,0)$ is a stable (unstable) fixed point of
the flow diagram when the ${\cal C}$ is positive (negative) near
$v=v_0$ (${\cal C}$ is assumed to be positive in
Fig.~\ref{fixed-points}). Flows running out of or into the fixed
points can be mutually interconnected and the way they are
interconnected will in general depend on the values of $v_g$ and
$q_g$ and affect the structure of the flow diagram.

To make our discussion concrete, we choose here a particular
hydrodynamic model.
We choose the coefficients in
Eq.~(\ref{VelocityTotalDeriv_Final})
as follows:
\begin{equation}
{\cal R} = \lambda,~~ {\cal A}={{\lambda \eta} \over{2\rho^3}},~~ {\cal D}=
{\lambda \over{6\rho^2}}, \label{afi}
\end{equation}
where $\eta \equiv d V_{\rm op}/d(\rho^{-1})$. As demonstrated in
Ref.~\cite{Lee01PRE}, this choice provides a macro-micro link
between the hydrodynamic model~[Eqs.~(\ref{VehicleConservation})
and (\ref{VelocityTotalDeriv_Final})] and the microscopic optimal
velocity model~\cite{Bando95PRE},
\begin{equation} \label{bando-model}
\ddot{y}_n=\lambda \left[ V_{\rm op}(\Delta y_n)-\dot{y}_n \right],
\end{equation}
where $y_n(t)$ represents the coordinate of the $n$-th vehicle
in a one-dimensional road and $\Delta y_n$ is the distance to
the preceding vehicle $y_{n+1}-y_n$.
But we remark that as far as steady state solutions are concerned,
the choice~(\ref{afi}) is just one {\it particular} option and
most results presented below are not sensitively dependent on it.
Results which are dependent on the choice will be stated so.
For the parameters, values in Ref.~\cite{Takaki98JPSJ} are
used: $\lambda=2$ (sec$^{-1}$),
\be \label{optimal-velocity}
V_{\rm op}(y)={v_{\rm mag} \over 2} \left[ \tanh {2(y-y_{\rm ref})
    \over y_{\rm width}}+c_{\rm ref} \right],
\ee $v_{\rm mag}=33.6$ (m/sec), $y_{\rm ref}=25.0$ (m), $y_{\rm
width}=23.3$ (m), and $c_{\rm ref}=0.913$. For this $c_{\rm ref}$,
the maximum value of $V_{\rm op}$ is $[(1+0.913)/2] v_{\rm mag}$,
which is slightly different from $v_{\rm mag}$.

\begin{figure}
\includegraphics[width=8.0cm, height=8.0cm]{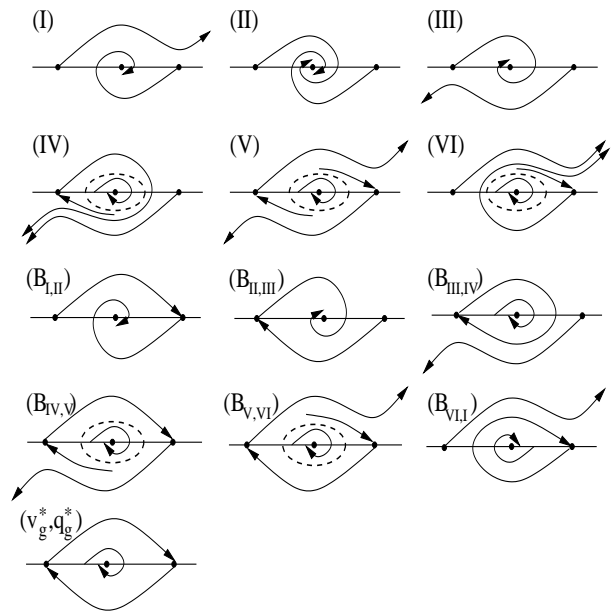}
\caption{
  Schematic drawings of the flow diagram in the phase
  space $(v,w)$. Depending on the values of $v_g$ and $q_g$, the flows
  running out of or into the fixed points are connected in different
  ways and as a result,
  the flow diagram acquires different topological structures.
  In each flow diagram, the fixed points
  $(v_{0,1,2},0)$ are marked by black circles. Emphasis is given to
  the flows running out of or into the fixed points in the range $v_1<v<v_2$.
}
\label{flow-diagrams}
\end{figure}

\begin{figure}
\includegraphics[width=8.0cm, height=6.0cm]{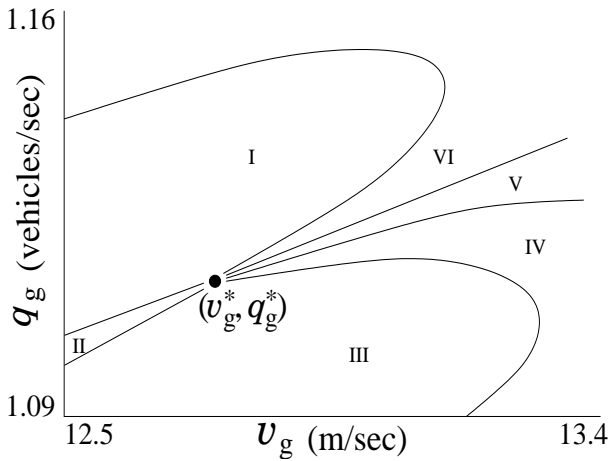}
\caption{The division of the parameter space $(v_g,q_g)$ based on
the topological structure of the flow diagram. } \label{sixdiv}
\end{figure}

For this hydrodynamic model, the resulting flow diagram is shown
for various values of $v_g$ and $q_g$ in Fig.~\ref{flow-diagrams}.
Note that depending on $v_g$ and $q_g$, trajectories departing
from the two saddle points $(v_{1,2},0)$ behave in different ways
and thus the flow diagrams acquire {\it topologically} different
structures. Since the structure of the flow diagram is closely
linked to characters of nonhomogeneous steady state solutions, it
will be meaningful to divide the plane $(v_g,q_g)$ according to
the flow diagram structures, which is given in Fig.~\ref{sixdiv}.
The $v_g$-$q_g$ plane is divided into six regions (region I, II,
$\cdots$, VI). The flow diagram {\it within} each region is
labeled accordingly in Fig.~\ref{flow-diagrams}. On the boundary
between two neighboring regions, (eg. boundary {\bf B}$_{\rm
I,II}$ between the region I and II) the flow diagram acquires
structures topologically different from those within the regions,
and on the special point $(v_g^*,q_g^*)$, where all six boundaries
join together, the flow diagram acquires a special structure still
different from all others. Steady state solutions contained in the
flow diagrams are presented in the next section.

\section{Steady state solutions}
\label{steady-state-solutions}

Out of all flow trajectories contained in the flow diagrams
(Fig.~\ref{flow-diagrams}),
only those trajectories that remain bounded both for $z\rightarrow \infty$
and $-\infty$ constitute physically meaningful steady state solutions.
Below we focus on those bounded trajectories.

\subsection{Saddle--Minimum solution}
\begin{figure}
\includegraphics[width=8.0cm, height=6.0cm]{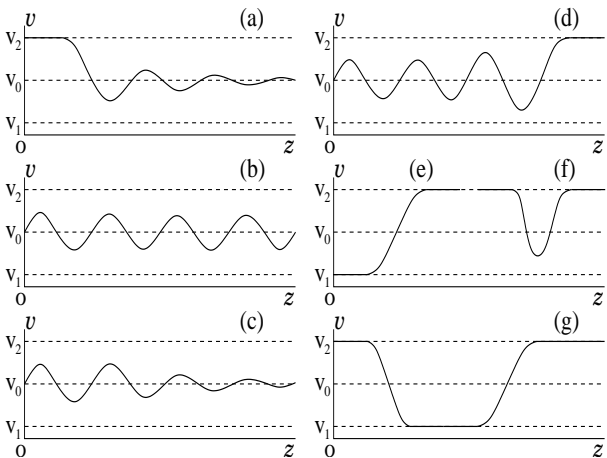}
\caption{Inhomogeneous steady state solutions of a hydrodynamic model.
}
\label{steady-states}
\end{figure}

For each $(v_g,q_q)$ in the region I and II, there exist a single
trajectory starting from the saddle point $(v_2,0)$ and converging
to the potential minimum point $(v_0,0)$ [see
Figs.~\ref{flow-diagrams}(I,II)]. This trajectory represents the
steady state solution in Fig.~\ref{steady-states}(a). Similarly
for each $(v_g,q_g)$ in the region II and III, there exist a
single trajectory starting from the other saddle point $(v_1,0)$
and converging to the potential minimum point $(v_0,0)$ [see
Figs.~\ref{flow-diagrams}(II,III)]. The steady state solution for
this trajectory is similar to that in Fig.~\ref{steady-states}(a)
except that the spatial profile approaches $v_1$ instead of $v_2$
as $z\rightarrow -\infty$. We call this type of steady state
solutions saddle-minimum solutions. We remark that the oscillation
near $v_0$ may or may not appear depending on the parameter
choice, which can be easily understood in the particle analogy; If
the particle motion near $v_0$ is underdamped/overdamped, the
convergence to $v_0$ in the saddle-minimum solution is
oscillatory/non-oscillatory. Expansion of Eq. (5) near $(v_0,0)$
shows that the motion near $v_0$ is overdamped if $\lambda>2$ and
underdamped if $\lambda<2$

A study~\cite{Berg01PRE} on the optimal velocity model revealed that
the microscopic model possesses so-called oscillatory traveling wave
solutions and monotonic wave solution. These solutions are identical
in their character to the saddle-minimum solution with the underdamped and
overdamped convergence to $v_0$.

\subsection{Limit-cycle solution}
For each $(v_g,q_g)$ in the region IV, V, or VI, there exists a
single trajectory, which encircles the potential minimum point
$(v_0,0)$ and makes a loop [see dashed curves in
Figs.~\ref{flow-diagrams}(IV,V,VI)]. In the language of nonlinear
dynamics, this type of flow is usually referred to as a
limit-cycle. For traffic flow language, the limit-cycle
corresponds to a steady state solution with periodic wave
[Fig.~\ref{steady-states}(b)]. The ``wavelength'' of the
limit-cycle solution increases and approaches infinite as
$(v_g,q_g)$ approaches $(v_g^*,q_g^*)$. We remark that for the
existence of the limit-cycle solution, it is crucial to have sign
alternation of ${\cal C}$ with $v$. If ${\cal C}$ were always
positive (negative), the single particle motion analogy indicates
that the total energy of the particle would monotonically decrease
(increase) with $z$, and thus the trajectory would be attracted to
(repelled away from) $v_0$, destroying the limit-cycle.

Note that since there is a limit-cycle solution for each $(v_g,q_g)$
within the region IV, V, and VI, there is an infinite number of limit-cycle
solutions,
each with different $(v_g,q_q)$.
This feature is very similar to the report of many stable nonhomogeneous states
in a revised car-following model in Ref.~\cite{Tomer00PRL}.
To our knowledge, it has not been realized previously that
hydrodynamic models also possess infinitely many limit-cycle solutions.

\subsection{Limit-cycle--Minimum solution}
The limit-cycle solutions are inevitably accompanied by still
different types of solutions. According to the flow diagrams in
Figs.~\ref{flow-diagrams}(IV,V,VI), all trajectories inside the
limit-cycle approach the minimum point $(v_0,0)$ as $z\rightarrow
\infty$, and the limit-cycle as $z\rightarrow -\infty$. We call
this type of steady state solutions limit-cycle--minimum solution.
Its profile is shown in Fig.~\ref{steady-states}(c). Since  there
are infinitely many limit-cycle solutions and the
limit-cycle--minimum solution is possible whenever the limit-cycle
is possible, there also exist infinitely many limit-cycle--minimum
solutions.

This solution is closely related with an interesting phenomenon
reported in previous studies of hydrodynamic models {\it with
on-ramps}~\cite{Helbing99PRL,Lee99PRE}: When a linearly unstable
but convectively stable homogeneous flow is generated in the
upstream part of an on-ramp, oscillatory flow is spontaneously
generated from the homogeneous region and propagate in the
upstream direction. The relation between this observation and the
limit-cycle--minimum solution can be established by noting that
the convectively stable homogeneous flow region corresponds to the
trivial solution $v(z)=v_0$ and the spontaneously generated
oscillatory flow to the limit-cycle solution. It is then clear
that the phenomenon in Refs.~\cite{Helbing99PRL,Lee99PRE} is
nothing but a manifestation of the limit-cycle--minimum solution.
Recent studies of a microscopic car-following
model~\cite{Mitarai99JPSJ} also reported oscillatory flow growing
out of a linearly unstable homogeneous region.

\subsection{Limit-cycle--Saddle solution}
Each limit-cycle solution accompanies still another type of
solutions. For each $(v_g,q_g)$ in the region IV and V, there
exists a single trajectory converging to the saddle point
$(v_1,0)$ as $z\rightarrow \infty$ and approaching the limit-cycle
as $z\rightarrow -\infty$~[Figs.~\ref{flow-diagrams}(IV,V)].
Similarly for each $(v_g,q_g)$ in the region V and VI, there exist
a single trajectory converging to the other saddle point $(v_2,0)$
as $z\rightarrow \infty$ and approaching the limit-cycle as
$z\rightarrow -\infty$~[Figs.~\ref{flow-diagrams}(V,VI)]. We call
this type of solutions limit-cycle--saddle solution. Its profile
in the $z$-space is shown in Fig.~\ref{steady-states}(d).
Similarly to the limit-cycle--minimum solutions, there exist
infinitely many limit-cycle--saddle solutions.

\subsection{Hetero saddle-saddle solution}
Still new types of bounded trajectories appear on the boundary
${\mathbf B}_{ij}$ between the region $i$ and $j$. On the
boundaries ${\mathbf B}_{\rm I,II}$ and ${\mathbf B}_{\rm IV,V}$,
there is a trajectory running from the saddle point $(v_1,0)$ and
converging to the other saddle point $(v_2,0)$
[Figs.~\ref{flow-diagrams}({\bf B}$_{\rm I,II}$),({\bf B}$_{\rm
IV,V}$)]. This trajectory amounts to the kink solution in
Fig.~\ref{steady-states}(e). Also on the boundaries ${\mathbf
B}_{\rm II,III}$ and ${\mathbf B}_{\rm V,VI}$, there is a
trajectory running from the {\it second} saddle point $(v_2,0)$
and converging to the {\it first} saddle point $(v_1,0)$
[Figs.~\ref{flow-diagrams}({\bf B}$_{\rm II,III}$),({\bf B}$_{\rm
V,VI}$)]. Since the roles of $(v_1,0)$ and $(v_2,0)$ have been
swapped, this trajectory amounts to the anti-kink solution [not
shown in Fig.~\ref{steady-states}]. In the nonlinear dynamics
language, these kinds of trajectories connecting two different
fixed points are called heteroclinic orbits. In this paper, we
will name this type of solutions hetero saddle-saddle solutions.
To be more specific, we will call the first (second) type of the
hetero saddle--saddle solutions ``upper'' (``lower'') hetero
saddle--saddle solutions since the trajectories appear in the
upper (lower) half of the flow diagram. Note that there exist
infinitely many hetero saddle--saddle solutions since this
solution is allowed for each point $(v_g,q_g)$ on the
abovementioned boundaries.

\subsection{Homo saddle--saddle solution}
For each $(v_g,q_g)$ on ${\mathbf B}_{\rm VI,I}$, there exists a
trajectory starting from the saddle point $(v_2,0)$ and returning
back to the {\it same} saddle point [Fig.~\ref{flow-diagrams}({\bf
B}$_{\rm VI,I}$)]. This trajectory represents the narrow cluster
solution [Fig.~\ref{steady-states}(f)]. For each $(v_g,q_g)$ on
${\mathbf B}_{\rm III,IV}$, on the other hand, there exists a
trajectory starting from the saddle point $(v_1,0)$ and returning
back to $(v_1,0)$ [Fig.~\ref{flow-diagrams}({\bf B}$_{\rm
III,IV}$)]. This trajectory represents the narrow anti-cluster
solution (not shown in Fig.~\ref{steady-states}). In the nonlinear
dynamics language, these kinds of trajectories are called
homoclinic orbits. In this paper, we will name this type of
solutions homo saddle-saddle solutions. To be more specific, we
will call the first (second) type of the homo saddle--saddle
solutions ``right'' (``left'') homo saddle--saddle solutions since
the trajectories involve the saddle point on the right (left) half
of the flow diagram.

\subsection{Wide cluster solution}
Figure~\ref{sixdiv} shows that all six regions and six boundaries
meet together at a {\it single} point $(v_g^*,q_g^*)$.
The flow diagram at this point has a special structure
[see Fig.~\ref{flow-diagrams}$(v_g^*,q_g^*)$]
that allow {\it smooth} connection between
flow diagrams of different topological structure
in different regions or boundaries.

This point is also special in the sense that
the flow running out of the fixed point $(v_1,0)$ reaches
the other fixed point $(v_2,0)$ and returns back to $(v_1,0)$.
The steady state solution corresponding to this flow is shown
in Fig.~\ref{steady-states}(g).
Note that since  the particle dynamics becomes infinitely slower as
a trajectory approaches the fixed points $(v_{1,2},0)$,
the cluster size of the steady state solution is infinitely large.
For this reason, we call this solution wide cluster solution.

The wide cluster solution is a limiting case of various solutions
mentioned above;
If $(v_g^*,q_g^*)$ is regarded as a limiting point of,
for example, the boundary $\mathbf{B}_{\rm III,IV}$ or $\mathbf{B}_{\rm VI,I}$,
the wide cluster solution is a homo saddle--saddle solution with
an infinite cluster size.
If $(v_g^*,q_g^*)$ is regarded as a limit point of the region VI, V, or VI,
the wider cluster solution is a limit-cycle solution with an infinite period.
Also if $(v_g^*,q_g^*)$ is regarded as the point where the borderlines
$\mathbf{B}_{\rm II,III}$
and $\mathbf{B}_{\rm I,II}$ join,
the wide cluster solution is a combined object of the upper hetero
saddle--saddle solution (kink)
and the lower hetero saddle--saddle solution (anti-kink).

The fact that the wide cluster solution is possible only at the
single point $(v_g^*,q_g^*)$ implies that the wide cluster
solution has the so-called {\it universal} characteristics; when a
given initial state of traffic evolves into the wide cluster
solution following the real traffic dynamics
[Eqs.~(\ref{VehicleConservation}) and
(\ref{VelocityTotalDeriv_Final})], characteristics such outflow
and the speed of the final traffic state are independent of the
initial state.

Analyses~\cite{Treiterer74Proceeding,Kerner96PRE} of empirical
traffic data revealed that various characteristics of the wide jam
are indeed universal. This empirical observation imposes a
constraint on traffic models and the universality of the
characteristics has been tested for traffic models. Although the
universality has been verified for a number of traffic
models~\cite{Komatsu95PRE,Kerner94PRE,Wada97CM,Helbing98PRL}, the
verification unfortunately has relied largely on repeated
numerical simulations and thus the verification itself is also
``empirical'' in some sense. One exceptional case is the analysis
in Ref.~\cite{Kerner97PRE}, where the singular perturbation theory
is used. This analysis is however considerably model-dependent and
thus it is not easy to perform the same analysis for other classes
of traffic models. Contrarily, in our approach, the new discovery
that the wide cluster solution can exist only at the single point
$(v_g^*,q_g^*)$ is quite meaningful since it explains clearly and
unambiguously why the characteristics of the wide cluster solution
should be universal.

We next address an interesting size dependence of the ``wide
cluster solution'' reported in Ref.~\cite{Kerner94PRE}; When
numerical simulations are performed with periodic boundary
conditions, the so-called universal characteristics are
found~\cite{Kerner94PRE,Wada97CM} to be not strictly universal but
depend on the system size. This size dependence is not compatible
with the statement of the universality given above. To resolve
this conflict, we first note that periodic boundary conditions
allow only those solutions whose periods are compatible with the
imposed period. Then the wide cluster solution, whose period is
infinite, can {\it not} be realized in such numerical simulations
with finite system size, and the solutions, which are interpreted
as the wide cluster solution in Ref.~\cite{Kerner94PRE,Wada97CM},
are, strictly speaking, limit-cycle solutions whose periods are
compatible with the imposed periodic boundary conditions. Then by
recalling that there are infinitely many limit-cycle solutions,
the size dependence is a very natural consequence and the conflict
is resolved. We remark that the dependence however should become
weaker as the period becomes longer since the limit cycle
solutions approach the wide cluster solution as their period
becomes longer.

\section{Discussion}
\label{discussion}

\subsection{Roles of topology: Generality}
As demonstrated in previous sections,
each steady state solution exists only in restricted regions of the
$v_g$-$q_g$ space where the flow diagram acquires certain
topological structures. This relation between the steady
state solutions and the topological structure of the flow diagram is
not a mere coincidence;
all steady state solutions presented in the preceding section
are {\it guaranteed} to exist by the topologies of the flow diagrams.
For example, when flows near the fixed point $(v_0,0)$ is attracted towards
$(v_0,0)$ while flows running out of the other fixed points
$(v_{1,2},0)$ are repelled away from $(v_0,0)$,
as in the regions (IV,V,VI),
there should exist the limit-cycle solution in those regions
(Poincar\'e-Bendixson theorem~\cite{Jackson89Book}).

This relation with the topology bears an interesting implication.
As demonstrated in many branches of physics,
physical objects, whose existence is closely related with a certain
topological structure of systems,
are {\it not fragile} and their existence does {\it not} depend
on quantitative details of the systems.
Vortices in type II superconductors are a well-known example.
It is then expected that
the steady state solutions presented in the preceding section
are not specific to the particular model examined but common to
many versions of hydrodynamic models.
For example, our results are not sensitive to the values of the
parameters in Eq.~(\ref{optimal-velocity}).

The generality due to the topology can be argued in the following
way as well. Let us consider an infinitesimal change in the
traffic model [in Eqs.~(\ref{afi}) and (\ref{optimal-velocity})
for example]. At $(v_g,q_g)$ located sufficiently interior of a
region in Fig.~\ref{sixdiv}, the topological structure of the flow
diagram will not be affected and thus all steady state solutions,
which are originally allowed in the region, are still allowed for
the modified traffic model. On the other hand, at $(v_g,q_g)$
located sufficiently close to a boundary in Fig.~\ref{sixdiv}, the
topological structure of the flow diagram may be modified to a new
structure. However the only possible new structure is the one at
just across the boundary. The net effect then amounts to a mere
shift of the boundary. Thus as far as the topology is concerned,
effects of the infinitesimal change in the traffic model is no
more than shifting the boundaries of the six regions by
infinitesimal amounts, and the existence of steady state solutions
in each region is {\it not} affected. Here we remark that despite
the shifts, all six boundaries should still meet at a {\it single}
point, whose coordinate may be slightly different from the
original $(v_g^*,q_g^*)$, though. Otherwise, continuous conversion
of one topological structure to another near $(v_g^*,q_g^*)$  is
not possible.

Lastly we discuss briefly one special kind of modifications that
shows negative-valued coefficient ${\cal C}$ [Eq.~(\ref{C&F})]
near the fixed point $(v_0,0)$. In the hydrodynamic model proposed
by Kerner {\it et al.}~\cite{Kerner93PRE}, ${\cal C}$ is indeed
negative when the parameters in the model are set to the values
suggested in Ref.~\cite{Lee99PRE}. For this case, the flow is {\it
repelled} away from $(v_0,0)$. Then the limit-cycle solution does
not appear any more in the regions IV, V, VI but appears instead
in the regions I, II, III. This shift is natural in view of the
Pointcar\'{e}-Bendixson theorem. However this negative ${\cal C}$
does not affect the very existence of the six regions because
these are determined according to the trajectories departing from
the two saddle point $(v_{1,2},0)$ as mentioned already. Thus our
approach focused on the bounded trajectories between the four
limiting behaviors (one limit-cycle and three fixed points) is
still valid. A more exotic kind of modifications are those, due to
which ${\cal C}$ changes its sign multiple times with $v$ so that
multiple limit-cycle solutions exist for given $v_g$ and $q_g$ in
certain regions of parameter space $(v_g,q_g)$. However this
possibility seems to be very unlikely since it requires
considerable fluctuations of ${\cal A}$ [see Eq.~(\ref{C&F})],
which is unphysical.

\subsection{Implications on universality conjecture}
A conjecture has been put forward by Herrmann and
Kerner~\cite{Herrmann98PA}; many traffic models with different
mathematical structures may belong to the same ``universality''
class in the sense that  they predict same traffic phenomena.
Although it is not clear yet to what extent the universality
conjecture is valid, there are indications that there indeed
exists close relationship between some traffic models. For
example, macroscopic hydrodynamic models  have been derived from a
microscopic car-following model via certain approximation
methods~\cite{Berg00PRE,Lee01PRE}, and good agreement between two
types of traffic models has been demonstrated via numerical
simulations~\cite{Helbing02MCM}. In particular, an exact
correspondence between two different types of microscopic models
has been established~\cite{Matsukidaira03PRL}. We note in passing
that the hydrodynamic model derived in Ref.~\cite{Berg00PRE}
however has an instability that arises from short-wave length
fluctuations that is {\it not} present in the original
car-following model. Such instability disappears when the
short-wave length fluctuations are properly regularized as in
Ref.~\cite{Lee01PRE}.

On the other hand, there exist reports which could not have been
reconciled with the universality conjecture.
For example, a recent study~\cite{Tomer00PRL} on a certain special
type of car-following model reported the existence of many limit-cycle
solutions (even without intrinsic inhomogeneities on roads such as
on-ramps).
To our knowledge, the limit-cycle solution had not been reported for
any other traffic models and thus it had been inferred that
the limit-cycle solution might be specific to the special model,
in clear contrast with the universality conjecture.
Our results however indicate that this inference is wrong
and reopen a possibility that the special car-following model
may also be closely related to other traffic models.
An evidence for this will follow in the next subsection.

By investigating topological structures of the flow diagrams,
we found from a {\it single} traffic model {\it seven} inhomogeneous
steady state solutions.
Although many of these solutions were already reported by earlier studies,
earlier reports were scattered over various different traffic models
and thus it was not clear whether a certain solution is specific to
certain traffic models or generic to a wide class of models.
Our analysis in the preceding subsection indicates
that the seven inhomogeneous steady state solutions are
generic to a wide class of traffic models.
Our results are thus valuable in view of the universality conjecture.

\subsection{Prevalence of limit-cycle solution in Ref.~\cite{Tomer00PRL}}
\begin{figure}
\includegraphics[width=8.0cm,height=6.0cm]{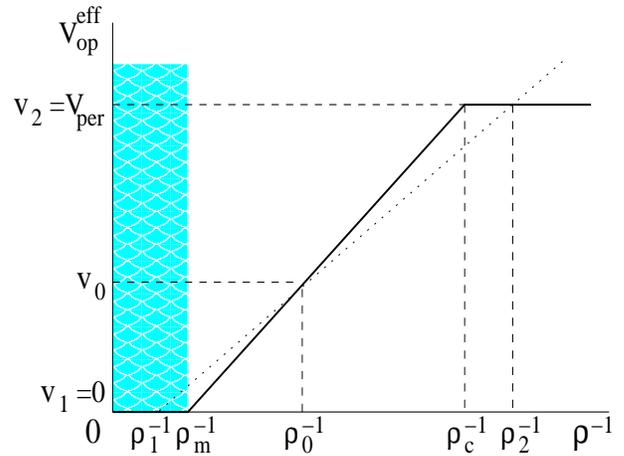}
\caption{The bold solid line is the effective optimal velocity
$V_{\rm op}^{\rm eff}$ due to the dynamics of
Eq.~(\ref{Tomer-model}). The dotted line on the other hand
describes the constant of motion line in Eq.~(\ref{Flux_in_vg}).
The hatched region, where no dynamics is allowed, is established
by the singular effect of the second term in
Eq.~(\ref{Tomer-model}). Three crossing points of the two lines
determine the $z$-independent solutions of Eq.~(\ref{Eqn_in_ss}).
Note that $\rho_c^{-1} = \rho_m^{-1} + T v_{\rm per}$. }
\label{sug_ov}
\end{figure}

As remarked above, most studies of traffic models failed to capture the
limit-cycle solution.
On the other hand, in the special car-following model studied in
Ref.~\cite{Tomer00PRL},
a wide class of initial conditions evolve to the limit-cycle solution.
A question arises naturally: what is special about the model in
Ref.~\cite{Tomer00PRL}?
The model is defined as follows,
\be
\label{Tomer-model}
\ddot{y}_n=A\left( 1-{\Delta y_n^0 \over \Delta y_n}
\right)-{Z^2(-\Delta \dot{y}_n) \over 2(\Delta
  y_n-\rho_m^{-1})}-kZ(\dot{y}_n-v_{\rm per}),
\ee where $Z(x)=(x+|x|)/2$, $A$ is a sensitivity parameter,
$\rho_m^{-1}$ is the minimal distance between consecutive
vehicles, $v_{\rm per}$ is the permitted velocity, $k$ is a
constant, and $\Delta y_n^0\equiv \dot{y}_n T+\rho_m^{-1}$. Here
$T$ is the safety time gap. When interpreted in terms of the
hydrodynamic model in Eq.~(\ref{VelocityTotalDeriv_Final}), the
first and third terms together define the effective optimal
velocity $V_{\rm op}^{\rm eff}(\rho^{-1})$, which is shown in
Fig.~\ref{sug_ov}. The role of the second term is to strictly
prevent the distance from being smaller than the minimum distance
$\rho_m^{-1}$ by establishing additional strong deceleration when
a vehicle is faster than its preceding one and their separation
approaches $\rho_m^{-1}$.

In view of the steady state analysis given in previous sections,
the optimal velocity profile in Fig.~\ref{sug_ov} is very special:
out of three solutions $v_1, v_2, v_3$ of $V_{\rm op}^{\rm
eff}((v+v_g)/q_g)=v$, which amount to the extremal points of the
potential $U^{\rm eff}$, $v_1$ is {\it dynamically forbidden}
since the vehicle spacing $\rho_1^{-1}$ related to $v_1$ via
Eq.~(\ref{Flux_in_vg}) is shorter than $\rho_m^{-1}$, which is
strongly prohibited by the second term in Eq.~(\ref{Tomer-model}).
Then steady state solutions, such as wide cluster solution, that
assume the accessibility to $(v_1,0)$ are not allowed any more and
the number of possible solutions reduces to 5 [saddle-minimum,
limit-cycle, limit-cycle--minimum, limit-cycle--saddle, and homo
saddle-saddle solutions]. Further reduction occurs when the
periodic boundary condition is imposed as in
Ref.~\cite{Tomer00PRL}. Then homo saddle-saddle and limit-cycle
solutions are the only possible solutions. Among these two, the
former is possible only when the average density satisfies
$\rho<(\rho_m^{-1}+T v_{\rm per})^{-1}$ because the unique
limiting behavior of that solution, of which density converges to
$\rho_2$, will govern the average density and $\rho_2 < \rho_c$.
Thus for the density range of $(\rho_m^{-1}+T v_{\rm
per})^{-1}<\rho<\rho_m$, the limit-cycle solution is the only
possible solution. Therefore the characteristics of the model in
Ref.~\cite{Tomer00PRL} is explained within the framework of the
hydrodynamic approach.

\subsection{Stability of solutions}
The rigorous stability analysis, whether each solution above is
realized through the dynamics in Eqs.~(\ref{VehicleConservation})
and (\ref{VelocityTotalDeriv_Final}), goes beyond the scope of
this work. In this section, we instead summarize what has been
known and also discuss implications of existing results. The
stability of the solution in Sec.~\ref{steady-state-solutions}~G
is well-established~\cite{Kerner93PRE,Kerner94PRE}. For the
solution in Sec.~\ref{steady-state-solutions}~C, its relation with
the oscillatory flow generated spontaneously out of a convectively
stable homogeneous flow, which was reported in previous numerical
simulations of hydrodynamic models with
on-ramps~\cite{Helbing99PRL,Lee99PRE}, seems to indicate that this
solution can be stable. Also the solution in
Sec.~\ref{steady-state-solutions}~B have been maintained stably in
our own numerical simulation of the hydrodynamic model that is
derived from the following microscopic model via the mapping rule
in Ref.~\cite{Lee01PRE}:
\begin{equation}
   \ddot y_n = \lambda \left [V_{\rm op}^{\rm eff} \left (
               \Delta y_n \right ) - \dot y_n \right ]
               + k{{\Delta \dot y_n}\over{\Delta y_n - \rho_m^{-1}}}
                 \ \Theta \left ( -\Delta \dot y_n \right ),
   \label{mod_OV}
\end{equation}
where $V_{\rm op}^{\rm eff}$ is the same one depicted in
Fig.~\ref{sug_ov}, $k$ is a constant, and $\Theta(x)$ is the
Heavyside function (1 for $x>0$ and 0 for $x<0$). Note that the
idea of the prevalence of the limit-cycle solution discussed in
Sec.~\ref{discussion}~C is simply reflected in Eq.~(\ref{mod_OV}).
These observations suggest that at least some steady state
solutions addressed in this work can be maintained stably. However
more systematic analysis is necessary to clarify the issue of the
stability.

\section{conclusion}
\label{conclusion}

Hydrodynamic traffic models are investigated by mapping them to the
problem of single particle motion. It is found that typical
hydrodynamic models possess seven different types of inhomogeneous
steady state solutions. Although these solutions were already reported
by earlier studies, earlier reports were scattered over various
different traffic models and it was not clear whether a certain
solution is specific to certain traffic models only or generic to a
wide class of models. Our result combined with the topology argument
indicates that the seven inhomogeneous steady state solutions should
be common to a wide class of traffic models.
Also the origin of the universal characteristics for the wide cluster
solution is clearly identified and the reason for the prevalence of
the limit-cycle solution in a previous report~\cite{Tomer00PRL} is provided.

\section*{Acknowledgments} This work is supported by the Korea Research
Foundation
(KRF 2000-015-DP0138).

\end{document}